 \documentclass[final,5p,times]{elsarticle}

 \usepackage{epsfig}

\usepackage{amssymb}
\usepackage{amsthm}

 \usepackage{lineno}

\journal{NIM A  RICAP-2013}

\begin{document}

\begin{frontmatter}



\title{Anisotropy in Cosmic rays from internal
transitions in neutron stars }


\author [1]{ M. \'Angeles P\'erez-Garc\'ia}
\author [2] { Kumiko Kotera}
\author [2] {Joseph Silk }

\address [1] {Department of Fundamental Physics and IUFFyM, University of Salamanca, Plaza de la Merced s/n 37008 Salamanca, Spain}
\address [2]{Institut d'Astrophysique de Paris, UMR 7095 - CNRS, Universit\'e Pierre $\&$ Marie Curie, 98 bis boulevard Arago, 75014, Paris, France}

\begin{abstract}
We discuss the possibility that some recently measured anisotropic cosmic ray components in the TeV-PeV energy range may be an indication of the ejection of a peculiar type of matter. We present a model where a neutron star internal transition with nuclear deconfinement of the quark content takes place. This catastrophic event may cause a mass ejection process seeding the insterstelar medium with  droplets of quark matter, so called nuclearites. Neutralization of these droplets in molecular clouds may drive the anisotropy since quasi-rectilinear trajectories are allowed. Complementary information from current experimental settings on earth or magnetic spectrometers on the ISS may shed light on this exotic form of  matter.
                
\end{abstract}


\end{frontmatter}


\section{Introduction}
\label{intro}

Galactic cosmic rays (CRs) have an energy spectrum showing characteristic features \cite{espec}. In particular their astroparticle nature (electrical charge, mass  or the ratio of them) has been pointed out as one of the key issues when trying to understand the souces where they may originate \cite{crdistri} and the emission processes itself. 
Additionally, this nature must determine, in turn, the mechanisms and feasibility to be accelerated to the high energies reported $E \sim 10^{20}$ eV. In this line there are some puzzling experimental measurements that are not yet fully understood. For example, several experiments have reported strong anisotropy measurements in the arrival direction distributions of Galactic CRs in the TeV to PeV energy range (Super-Kamiokande, Tibet III, Milagro, ARGO-YBJ, and IceCube \cite{neutrino_exp,Abbasi11}). The data reveal the presence of large scale anisotropies of amplitude $\sim 0.1\%$. Smaller scale anisotropies of size $\sim10^\circ-30^\circ$ are also detected with amplitude a factor of a few lower. Milagro has reported the detection at significance $>12\sigma$ of two {\it hotspots} (regions with enhanced CR intensity) with amplitude $\approx 10^{-4}$, at a median energy of $1\,$TeV. ARGO-YBJ report similar excesses. IceCube observes localized regions of angular scale $\sim 15^\circ$ of excess and deficit in CR flux with significance $\sim5\sigma$ around a median energy of 20\,TeV~\cite{Abbasi11}. 

The large scale anisotropy could be naturally explained by the diffusive transport of CRs within the Galactic magnetic fields \cite{large_scale_theories,previous_works}. Due to the charged nature of CRs at energies in TeV-PeV range their propagation is described by a gyroradius  (Larmor radius) given by 
\begin{equation}
r_{\rm L} \approx \frac{E}{ZeB} \sim 1.08\,{\rm pc}\,Z^{-1} \left( \frac{E}{1\,{\rm PeV}}\right) \left(\frac{B} {1\,\mu{\rm G}}\right)^{-1},
\end{equation}
where $Z$ is the charge of the particle in units of the electron charge $e$  and the magnetic field strength of the Galaxy is assumed to be $B=1\,\mu$G (see \cite{Han08} for a review). For particles with $r_{\rm L}\ll l_{\rm c}$, where $l_{\rm c}=10-100\,\rm pc$ is the coherence length of the Galactic magnetic field (e.g., \cite{Han08}), the propagation will be totally diffusive over a  distance $>l_{\rm c}$. A number of previous works \cite{previous_works} have attempted  to explain these phenomena invoking several mechanisms, however the situation remains largely uncertain. 

In this contribution based in \cite{kotera} we develop the possibility that the measured hotspots in the skymap are a manifestation of the peculiar nature of CRs. We propose that quark matter lumps, so-called {\it strangelets} or {\it nuclearites}, could be produced in the mass ejection process taking place in the nuclear deconfinement transition of a regular neutron star (NS)  to a quark star (QS). This possibility has been proposed long ago \cite{itoh} \cite{witten84} and revisited in later works \cite{ouyed02} \cite{cr_acc}. Recently, new ideas concerning the triggering due to presence of a internal energy release from a dark matter component in a   sort of {\it Trojan horse mechanism} have been considered \cite{Perez-Garcia10}\cite{Perez-Garcia11}. Additional sources for these droplets may arise from high-density environments of merger events \cite{merger} \cite{and}.

These slightly positively charged lumps of quark matter may suffer a diffusive trajectory and if molecular clouds (MC) are near the sources this may drive the anisotropy. Possible processes in the cloud include electron capture, decay or even spallation \cite{berger}. In this way, for example, a change in the droplet incident state of charge, as a consequence of the interaction with the MC may neutralize the lump or it could decay with some fragments likely to be neutralized in the cloud.

\section{Strangelets} 
These lumps would be formed after the metastable $ud$ matter decays by weak interaction, $u+d\rightarrow u+s$, to form more stable $uds$ matter \cite{itoh} \cite{witten84}.  They are expected to be highly bound $m_A\lesssim Am_N$ .  Typical values of strangelet binding energy are currently uncertain but supposed to be $E/A \sim \rm MeV-GeV$ energies. Either on earth (accelerators) or on the ISS (with the AMS-02 spectrometer ) direct searches are being conducted to experimentally detect this (so far) elusive type of astroparticles.

The lowest energy state in a strangelet is not subject to the constraint of being  neutral, and therefore it is energetically allowed to have stable $Z/A>0$, $Z/A \ll 1$ massive strangelets \cite{Madsen}, where  $A$ is the baryonic number. There is, however, a  constraint on the minimum value of $A\sim 10-600$ \cite{Wilk96}. Several models of strangelets exist that lead to various $Z/A$ dependencies. For example, for ordinary strangelets, $Z\sim A^{1/3}$, while for CFL (color-flavour-locked) strangelets $Z\simeq 0.3 A^{2/3}$ \cite{Madsen}. Even smaller charge-to-mass ratios are allowed $Z/A\sim10^{-7}\div 10^{-2}$. Regarding charge, experiments such as CREAM and AMS-02 will have the ability to perform a direct measurement, and infer estimates of $Z/A$ \cite{arruda} with the RICH instrument. 

If strangelets were responsible for the  observed hotspots, they should produce detectable air-showers. There is a general belief that these should manifest as slowly moving droplets providing an enhanced photon production as they cross the water based telescopes \cite{ant}. In turn, this is possible if the kinetic energy per nucleon content, $K_{N}$, satisfies $K_N=K_{\rm tot}/A>1\,$GeV. Measurements indicate a total kinetic energy of particles in hotspots, $K_{\rm tot}\sim E \sim$\,TeV-PeV, which implies $A\lesssim 10^2-10^4$.

\begin{figure}[hbtp]
\begin{center}
\includegraphics [angle=-90, scale=0.75] {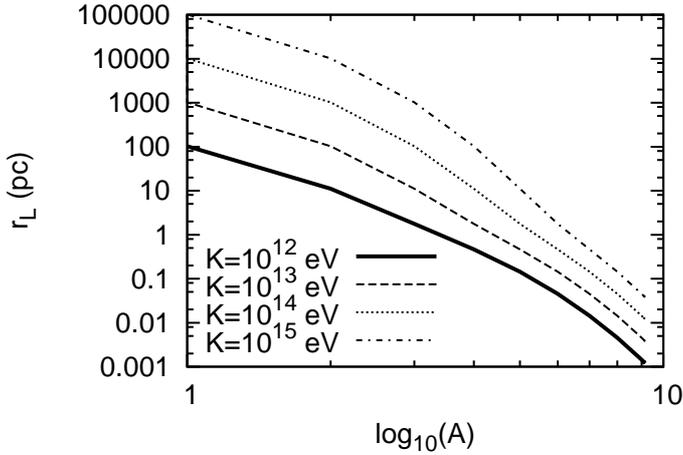}
\caption{Larmor radius as a function of $A$ for strangelets with $Z/A\sim 10^{-6}$ and energies in the $\sim$TeV-PeV range.}
\label{fig:rl}
\end{center}
\end{figure}

Fig.~\ref{fig:rl} represents the droplet Larmor radius as a function of the baryonic number $A$ for strangelets with $Z/A\sim 10^{-6}$ and kinetic energy contours  $K=10^{12}, 10^{13}, 10^{14}, 10^{15}$ eV. Typically a diffusive behaviour is expected, since it is required that $Z>1$. If, instead, charge could be fractionary then quasi-rectilinear regimes would be possible. 

\section{Astroparticle sources}

Neutron stars have been suggested as possible sources of injection of strangelets \cite{strangelet_accelerators}. Strangelets could be produced for instance in the course of a NS to QS transition \cite{strange_stars}. In such events, a fraction $f_{\rm ej}$ of the gravitational energy released can be injected into the expelled outer crust, leading to total kinetic energies $E_{\rm ej}\sim 4\times10^{50}(f_{\rm ej}/10^{-3})\,$ erg for standard NS mass and radius. The Lorentz factor $\Gamma$ of the ejected mass can be of order 
\begin{equation}
\Gamma\sim 22\,\left(\frac{f_{\rm ej}}{10^{-3}}\right)\left(\frac{12\,{\rm km}}{R_{\rm *}}\right)\left(\frac{M_{\rm *}} {1.5M_\odot}\right)^2 \left(\frac{10^{-5}M_\odot}{M_{\rm ej}}\right),
\end{equation}
for NS mass $M_{\rm *}$, radius $R_{\rm *}$, and ejected mass $M_{\rm ej}$ \cite{Perez-Garcia12}. Particles of mass number $A$ could then gain energies of order $K_{\rm acc}\sim 21\,(A/10^3)(\Gamma/22)\,{\rm TeV}$, the typical energy observed in hotspots.
It has been shown \cite{Perez-Garcia12} that typical ejection fractions depend on the characteristics of the transition and, in particular, this may generate a multi-wavelength signal to help discriminate this catastrophic astrophysical event. A short hard $E_{\gamma}\gtrsim 100$ keV spike in gamma rays is predicted to appear.
\begin{figure}[hbtp]
\begin{center}
\includegraphics [scale=0.4] {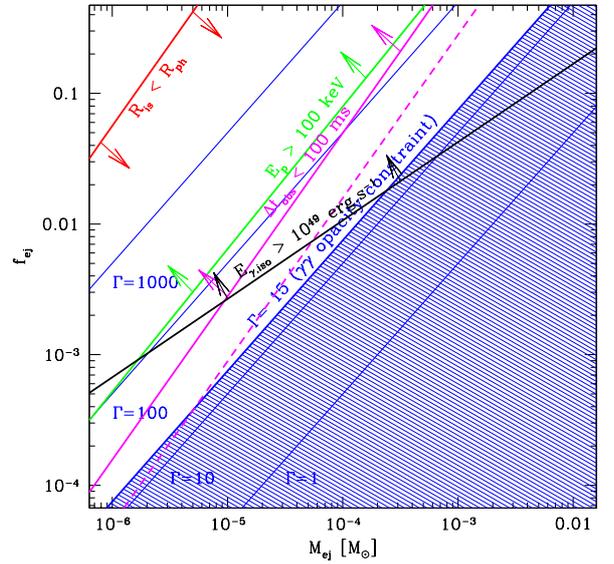}
\caption{Efficiency of the energy injection in the crust $f_\mathrm{ej}$ in a NS transition versus ejected outer crust $M_\mathrm{ej}$  from \cite{Perez-Garcia12}. See text for details.}
\label{fig:grbnsqs}
\end{center}
\end{figure}

In Fig. \ref{fig:grbnsqs} we plot the efficiency of the energy injection in the crust $f_\mathrm{ej}$ in a NS transition versus ejected outer crust $M_\mathrm{ej}$ as appears in \cite{Perez-Garcia12}.  Lines of constant Lorentz factor $\Gamma$ are plotted in blue for $\Gamma=1$ (non relativistic limit), $10$, $15$, $100$ and $1000$. 

The limit $\Gamma\simeq 15$ obtained from the compactness argument limits the forbidden shaded region where a gamma ray burst emission is not visible from kinematical constraints. Time and peak energy, isotropic equivalent gamma-ray energy observability limits are shown in magenta, green and black respectively. Astrophysical radii constraints are depicted in red.

Accelerated strangelets may experience energy losses by interacting with the radiation field close to the NS, and with the baryonic and radiative backgrounds of the supernova (SN) envelope. Refs.~\cite{cosmic_rays_pulsars} concluded that there is room for the escape of accelerated particles.
Besides, if the ejection happens when there are no SN envelope (since the NS may have traveled far after the initial birth) in a transition of a NS to a QS then ejection may be more efficient.

\section{Interaction in the MC and induced anisotropies}

Charged strangelets will have a diffusive trajectory due to the magnetized interstellar medium (ISM). Typical timescales are
\begin{equation}
\Delta t=\frac{d_{\rm s}^2}{2D}\sim 6\times 10^5\, Z^{1/3}\left(\frac{d_{\rm s}}{1\,{\rm kpc}}\right)^2 \left(\frac{E}{{\rm 20\,TeV}}\right)^{-1/3}\,{\rm yr},
\end{equation}
where $d_{\rm s}$ is the rectilinear distance to the source and the diffusion coefficient is $D(E) = 1.33\times10^{28}H_{\rm kpc}[E/(3Z\,{\rm GeV})]^{1/3}$\,cm$^2\,$s$^{-1}$, with $H_{\rm kpc}\equiv H/(1\,{\rm kpc})$ the height of the Galactic halo \cite{Blasi12_1}. 
The ionization and the spallation timescales in the ISM (of average density $n_{\rm ISM}=0.5\,{\rm cm}^{-3}$) read respectively $\tau_{\rm ion}\sim 7\times10^{12}\,Z^{-2}\left(\frac{E}{{\rm 20\,TeV}}\right)\,$yr, and $\tau_{\rm spall}\sim 4\times 10^5\,(A/10^3)^{-2/3}\,\left(\frac{n_{\rm ISM}}{ 0.5\, {\rm cm^{-3}}}\right)^{-1}\,$yr \cite{Madsen}, implying that spallation should affect particles only mildly during their flight from sources located within 1\,kpc. Let us consider that in our proposed model \cite{Perez-Garcia10} \cite{kotera} NS transitions are more likely to happen in old objects where dark matter accretion may have sufficient time to drive the conversion.

In the MC the typical radius in the Galaxy is $R_{\rm MC}\sim 20-50\,$pc, and their average density $n_{\rm MC}\sim10^{2-4}\,{\rm cm}^{-3}$. Cores are the inner more dense regions $0.1$ pc where $n_{\rm {core}}\sim10^{5-6}\,{\rm cm}^{-3}$ and fields $B\sim 100 \,\mu {\rm G}\, (n/10^{4}\,{\rm cm}^{-3})$. In MC the spallation fraction can exceed unity, reaching $r_{\rm spall}=\tau_{\rm esc}/\tau_{\rm spall}$. $r_{\rm spall}\sim 7.5\,Z^{1/3}\left(\frac{R_{\rm MC}}{25\,{\rm pc}}\right)\left(\frac{n_{\rm MC}}{10^3\,{\rm cm}^{-3}}\right)\left(\frac{A}{10^3}\right)^{2/3}$, with $\tau_{\rm esc}$ the diffusion time of strangelets in the cloud. The electron capture rate for strangelets in clouds with free electron density $\sim \eta_e n_{\rm MC}$ (with $\eta_e<<1$) is of order $r_{\rm ion}\sim 10^{-5}Z^{7/3}\eta_e (n_{\rm MC}/10^3\,{\rm cm}^{-3})$. As strangelets are predicted to be more bound than standard nuclei, these estimates can be viewed as upper limits for spallation. Let us note that for a large $Z/A$ lump of quark matter Z may be large and the amount of ionization may not be negligible causing scintillation effects \cite{scint}.
Additionally, for electron capture, it is possible that the large size of strangelets dominates the effects of the charge, implying a scaling in $R^2 \sim A^{2/3}$, and the rates quoted here can be viewed as a lower limit. 

\subsection{Neutralization} 

One could expect that a fraction of strangelets undergoing  electron capture (similar to that quoted for regular ions \cite{Padovani09}) or even spallation may generate charge {\it neutral} secondaries. Although most of the strangeness carrying lumps of quark matter is charged the work of \cite{Madsen} suggests that a tiny parameter space exists where spallation could lead to bound neutral strangelets.

Neutral strangelets could then propagate rectilinearly to the earth and produce a {\it hotspot} in the sky of the angular size of the MC, $\theta_{\rm MC}\sim 14^\circ\,(R_{\rm MC}/25\,{\rm pc})(d_{\rm MC}/200\,{\rm pc})^{-1}$, with $d_{\rm MC}$ the distance of the MC to the observer. Note that this corresponds roughly to the size of the observed hotspots. 

The excess signal in a solid angle $<\Omega$ around one source can be defined as the following signal-to-noise ratio:
$\sigma_{<\Omega} = N_{\rm s,<\Omega}/(N_{\rm iso,<\Omega})^{1/2}$,
where $N_{\rm s,<\Omega}=
{L_{\rm MC}}A(\alpha,\delta){4\pi d_{\rm s-MC}^2\Omega E}^{-1}
$ 
indicates the number of events expected in a solid angle $<\Omega$ from a source and $N_{\rm iso,<\Omega}=  E{J_{\rm iso,sr}}A(\alpha,\delta)$ the corresponding number of events expected for an isotropic background. For a MC located at coordinates $(\alpha,\delta)$, at distance $d_{\rm MC}$, and separated by $d_{\rm s-MC}$ from the source, the signal at energy $E$ can then be estimated as: 
\begin{equation}
\label{eq:sigma}
 \sigma(E)=\frac{\eta}{E^{3/2}}\left[1+\frac{ d_{\rm s-MC}^4c^2}{4D^2R_{\rm MC}^2}\right]^{-1}\frac{E_{\rm ej}}{\Delta t}\frac{A(\alpha,
\delta)^{1/2}}{4\pi d_{\rm MC}^2\Omega J_{\rm iso,sr}^{1/2}}\ ,
\end{equation}
where $A(\alpha,\delta)$ [in m$^{2}$~s~sr] is the exposure of an experiment in the direction $(\alpha,\delta)$, $J_{\rm iso,sr}(E)$ is the observed cosmic ray flux at energy $E$, per steradian, and $\Delta t$ is the diffusion time for particles to travel over a distance $\min(R_{\rm MC},2R_{\rm MC}+d_{\rm MC})$. The factor $\eta \sim \eta_s f_{\rm neutr}$ accounts for strangelet production rate at the source, and the neutralization rate in the MC.

For a source located inside the MC, the luminosity in neutral strangelets radiated by the MC at $E=20\,$TeV is of order $L_{\rm MC}=E_{\rm MC}/\Delta t\sim 3.5\times10^{40}\,\eta Z^{-1/3}(R_{\rm MC}/25\,{\rm pc})^{-2}\,{\rm erg/s}$. Cosmic ray measurements indicate $J_{\rm iso,sr}(20\,{\rm TeV})\sim5\times 10^{-17}$\,eV$^{-1}$~{s}$^{-1}$~m$^{-2}$ sr$^{-1}$, and we chose an exposure of $A(\alpha,\delta)= 10^{13}\,$m$^2\,$s\,sr, roughly corresponding to the Milagro exposure  at $20\,$TeV, over 7 years of operation.

\begin{figure}[tp]
\begin{center} 
\includegraphics[width=\columnwidth]{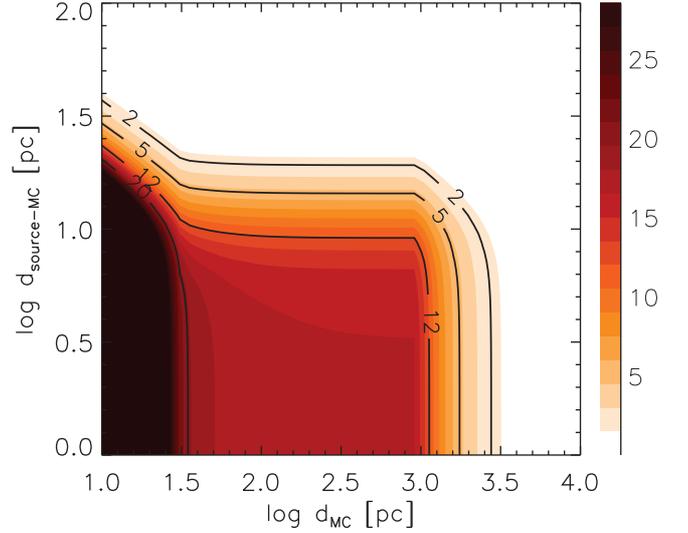}
\caption{Particle excess significance $\sigma$ (Eq.~\ref{eq:sigma}), as would be observed by Milagro with 7 years of data, at $E=20\,$TeV, as a function of  the distance of the MC to the earth, $d_{\rm MC}$, and the distance between the source and the MC, $d_{\rm s-MC}$ from \cite{kotera}.  }
\label{fig:sigma}
\end{center}
\end{figure}

In our calculation, we consider a production rate $\eta_s\sim 0.01$ (see Fig. \ref{fig:grbnsqs}) while the neutralization must be related to the degree of ionization present in the MC. Typically $\xi_H\sim \rm few 10^{-15} s^{-1}$ and since diffusion times in this regime are $\Delta t\sim 100\, \rm yr\sim 10^9$, then  $f_{\rm neutr}\sim 10^{-6}$. This leads to an efficiency factor $\eta \sim \rm few 10^{-8}$. 

We find that efficiencies in the range of $10^{-8}<\eta<10^{-7}$ lead to reasonable values in terms of $\sigma$ whatever the relative location of the source and the MC, and the distance to the MC. From Eq.~\ref{eq:sigma}, one can infer the strong dependency of $\sigma$ on the distance between the source and the MC: $\sigma\propto d_{\rm s-MC}^{-4}$, when the source is at the border of the MC.  , Fig.~\ref{fig:sigma} depicts a case for strangelets with $Z=1$, $A=10^3$ and a MC of radius $R_{\rm MC}=25\,$pc and a source of luminosity $L_{\rm MC}=\eta10^{40}\,$erg/s with an efficiency factor $\eta = 5\times 10^{-8}$. The value of $\sigma$ is relatively constant as long as the source is at a relatively central position inside the MC. This range of $\eta$ thus implies that only MC within $1-2\,$kpc, and only sources {\it located inside the MC} can produce a significant hotspot (note also that local MCs are found beyond $d_{\rm MC}\gtrsim 70\,$pc). 

Most observed hotspots could be produced by MCs in the Gould Belt (a star forming region concentrating many MCs, that forms a ring at a distance from the Sun of $\sim 0.7-2\,$kpc), at the location where NS-QS transitions may have occurred. Interestingly, the Milagro hotspot labelled ``Region A" \cite{neutrino_exp} lies in the direction of the Taurus Molecular Cloud, the nearest star formation region located at 140\,pc, and that covers $\sim 100\,{\rm deg}^2$ in the sky \cite{clouds}. ``Region 1" of IceCube \cite{Abbasi11} is also in the direction of a remarkable MC: the Vela Molecular Ridge, located at $0.7-2\,$kpc distance, of size $\sim 15^\circ$ in sky \cite{clouds}.

We discussed the possibility that strangelets accelerated in nearby NS-QS transitions, and then becoming neutral in molecular clouds, could explain the small-scale anisotropies observed by several experiments at TeV-PeV energies. \\

We thank the COMPSTAR and MULTIDARK projects, Spanish MICINN projects FIS-2009-07238 and FIS2012-30926. M.A.P.G would like to thank the kind hospitality of IAP where part of this work was developed. K.K. acknowledges support from PNHE.







\begin{thebibliography}{9}
%
\bibitem{espec} A. Haungs et al., Rep. Prog. Phys. 66 1145 (2003)
\bibitem{crdistri} A. M. Hillas, Annual review of astronomy and astrophysics. Volume 22. Palo Alto, CA, p. 425 (1984) 
\bibitem{neutrino_exp} G. Guillian et al., PRD 75, 062003 (2007); M. Amenomori et al., Science 314, 439 (2006); 
A. A. Abdo et al., PRL 101, 221101 (2008); A. A. Abdo, et al., ApJ 698, 2121 (2009); S. Vernetto, Z. Guglielmotto, J. L. Zhang, and for the ARGO-YBJ Collaboration, ArXiv: 0907.4615 (2009); R. U. Abbasi et al., PRL 104, 161101 (2010)
\bibitem{Abbasi11} R. Abbasi, et al., ApJ 740, 16 (2011)
\bibitem{large_scale_theories} A. D. Erlykin and A. W. Wolfendale, Astroparticle 25, 183 (2006);  P. Blasi and E. Amato, JCAP 1, 11 (2012)
\bibitem{previous_works} L. O. Drury and F. A. Aharonian, Astroparticle Phys. 29, 420 (2008); 
M. Salvati and B. Sacco, A\&A 485, 527 (2008); G. Giacinti and G. Sigl (2011), ArXiv:1111.2536
\bibitem{Han08} J. L. Han, Nuclear Physics B Proc. Suppl. 175, 62 (2008)
\bibitem{kotera} K. Kotera, M. A. Perez-Garcia, J. Silk, arXiv:1303.1186, Phys. Lett. B 725, 196 (2013).

\bibitem{itoh} N. Itoh,  Prog. Theor. Phys., 44, 291(1970) 
\bibitem{witten84} E. Witten, Phys. Rev. D, 30, 272 (1984)
\bibitem{ouyed02} R. Ouyed, J. Dey, and M. Dey, A\&A, 390,  L39 (2002)
\bibitem{cr_acc} G. A. Medina and J. E. Horvath, ApJ 464, 354 (1996)
\bibitem{Perez-Garcia10} M. A. Perez-Garcia, J. Silk, and J. R. Stone, PRL 105, 141101 (2010)
\bibitem{Perez-Garcia11} M. A. Perez-Garcia and J. Silk, ArXiv:1111.2275
\bibitem{merger} R. Oechslin, K. Uryu, G. Poghosyan, F. K. Thielemann, Mon. Not. Roy. Astron. Soc., 349, 1469, (2004)
\bibitem{and} A. Bauswein, H.-Th. Janka, R. Oechslin  et al., Phys. Rev. Lett. 103,  011101 (2009)
\bibitem{berger} M.S. Berger, R. L. Jaffe, Phys. Rev. C 35, 213  (1987)
\bibitem{Madsen}  J. Madsen, PRL 85, 4687 (2000); J. Madsen, PRL 87, 172003 (2001); J. Madsen, PRD 71, 014026 (2005); J. Madsen, arXiv:0612740;  J. Madsen, arXiv:0512512
\bibitem{Wilk96} G. Wilk and Z. Wlodarczyk, Journal of Physics G Nuclear Physics 22, L105 (1996)
\bibitem{arruda} L. Arruda, F. Barao, R. Pereira,  arxiv: 0710.0993v1
\bibitem{ant} G. Giacomeli, Antares collab, arXiv:1211.5516
\bibitem{strangelet_accelerators} J. Madsen, PRD 71, 014026 (2005); K. S. Cheng and V. V. Usov, PRD 74, 127303 (2006)
\bibitem{strange_stars} C. Alcock, E. Farhi, and A. Olinto, ApJ 310, 261 (1986); C. Alcock and A. Olinto, ARNPS 38, 161 (1988)
\bibitem{Perez-Garcia12} M. A. Perez-Garcia, F. Daigne, and J. Silk, ApJ 768, 145 (2013)
\bibitem{cosmic_rays_pulsars} W. Bednarek and R. J. Protheroe, PRL 79, 2616 (1997); P. Blasi, R. I. Epstein, and A. V. Olinto, ApJ Letters 533, L123 (2000); J. Arons, ApJ 589, 871 (2003); K. Kotera, PRD 84  (2011) 023002, K. Fang, K. Kotera, and A. V. Olinto, ApJ 750, 118 (2012)
\bibitem{Blasi12_1} P. Blasi and E. Amato, JCAP 1, 10 (2012)  
\bibitem{scint} M. A. Perez-Garcia, J. Silk,U. L. Pen,  arXiv:1304.8116.
\bibitem{Padovani09} M. Padovani, D. Galli, and A. E. Glassgold, A\&A 501, 619 (2009).
\bibitem{clouds} G. Narayanan, M. H. Heyer, C. Brunt, P. F. Goldsmith, R. Snell, and D. Li, ApJ S. 177, 341 (2008); D. C. Murphy and J. May, A\&A 247, 202 (1991)

\end{thebibliography}







\end{document}